\documentclass[journal]{IEEEtran}
\usepackage{textcomp}
\usepackage{xcolor}
\def\BibTeX{{\rm B\kern-.05em{\sc i\kern-.025em b}\kern-.08em
    T\kern-.1667em\lower.7ex\hbox{E}\kern-.125emX}}

\usepackage{booktabs}
\usepackage{epsfig,rotating,setspace,latexsym,epsf,bm}
\usepackage{cite,color, hyperref}
\usepackage{amsmath,amssymb,amsfonts}
\usepackage{graphicx}
\usepackage{graphics}
\usepackage{subcaption}
\usepackage{balance}
\usepackage{comment}
\usepackage{courier}
\usepackage{algpseudocode}
\usepackage[ruled,linesnumbered]{algorithm2e}
\usepackage{multirow}
\usepackage{multicol}
\usepackage{makecell}
\usepackage{float}
\usepackage{tikz}

\usepackage{caption}
\usepackage{tikz}

\begin{document}

\title{System-Level Design Space Exploration for High-Level Synthesis under End-to-End Latency Constraints}

\author{Yuchao Liao,~\IEEEmembership{Member,~IEEE}, Tosiron Adegbija,~\IEEEmembership{Senior Member,~IEEE}, and Roman Lysecky,~\IEEEmembership{Senior Member,~IEEE} 
\thanks{The authors are with the Department of Electrical and Computer Engineering, The University of Arizona, USA, email: \{yuchaoliao, tosiron, rlysecky\}@arizona.edu.}
\thanks{This work is partly supported by the National Science Foundation under grant CNS-1844952.}
}

\maketitle

\begin{abstract}
Many modern embedded systems have end-to-end (EtoE) latency constraints that necessitate precise timing to ensure high reliability and functional correctness. The combination of High-Level Synthesis (HLS) and Design Space Exploration (DSE) enables the rapid generation of embedded systems using various constraints/directives to find Pareto-optimal configurations. Current HLS DSE approaches often address latency by focusing on individual components, without considering the EtoE latency during the system-level optimization process. However, to truly optimize the system under EtoE latency, we need a holistic approach that analyzes individual system components' timing constraints in the context of how the different components interact and impact the overall design. This paper presents a novel system-level HLS DSE approach, called \textit{EtoE-DSE}, that accommodates EtoE latency and variable timing constraints for complex multi-component application-specific embedded systems. EtoE-DSE employs a latency estimation model and a pathfinding algorithm to identify and estimate the EtoE latency for paths between any endpoints. It also uses a frequency-based segmentation process to segment and prune the design space, alongside a latency-constrained optimization algorithm for efficiently and accurately exploring the system-level design space. We evaluate our approach using a real-world use case of an autonomous driving subsystem compared to the state-of-the-art in HLS DSE. We show that our approach yields substantially better optimization results than prior DSE approaches, improving the quality of results by up to 89.26\%, while efficiently identifying Pareto-optimal configurations in terms of energy and area.
\end{abstract}

\section{Introduction}
Embedded systems are transforming various industries, enabling breakthroughs in medical devices, autonomous vehicles, agriculture, and the growing Internet of Things (IoT). These systems often necessitate the use of application-specific hardware to fulfill stringent timing requirements like variable timing and end-to-end (EtoE) latency constraints while operating within resource limitations, such as area and energy. %The design and optimization of such application-specific embedded systems demand careful attention to precise timing constraints. This is especially crucial for human-centric use cases, like wearable healthcare devices or autonomous driving systems, where achieving accurate real-time sensing, computation, and communication within stringent area and energy constraints poses a critical challenge\cite{LIAOHHLS2022}. 
High-Level Synthesis (HLS) has emerged as a powerful solution, allowing designers to automatically generate hardware implementations from high-level languages like C/C++/SystemC, rather than relying on low-level hardware description languages (HDLs) such as Verilog or VHDL. This approach significantly increases productivity and reduces development times and costs, making HLS a compelling choice for hardware design, especially for human-centric applications where precise real-time sensing, computation, and communication within tight constraints are paramount. Studies have shown that HLS-style development can reduce designer effort by up to 6 times compared to traditional methods \cite{lahti2018we}.

HLS tools empower designers to customize synthesis directives, such as target frequencies, resource bindings, and loop unrolling to meet the unique requirements of application-specific embedded systems. However, the abundance of synthesis directives, along with interrelated objectives like energy and area, has led to a supra-linear expansion of the design space, making exhaustive exploration infeasible. Design Space Exploration (DSE) techniques, often employing heuristic algorithms due to the complexity of the problem, are commonly used to identify Pareto-optimal configurations \cite{LIAOHHLS2022,GOSWAMI2023116,SohrabizadehAutoDSE2021,YeAutoscaleDSE2023}. 

The inclusion of precise timing constraints in complex multi-component systems further exacerbates the design space complexity. While models like the periodic state machine (PSM) \cite{kopetz07_PSM,LIAOHHLS2022} with a fixed period have been used to incorporate precise timing specifications, current HLS tools \cite{vivadoHLS2023,SmartHLS2023,IntelHLS2023,ferrandi2021bambu,LIAOHHLS2022} and DSE methods offer limited support for variable timing constraints, especially for embedded systems requiring system-level and component-level EtoE latency constraints across diverse components. A holistic approach that considers both individual component timing and inter-component interactions is crucial to effectively optimize EtoE latency in multi-component systems. 

To motivate the need for holistic system-level DSE, consider an autonomous vehicle's object detection system. Traditional approaches might independently optimize camera capture speed, image processing, and output generation. However, this isolated optimization neglects the components' interactions and overall system impact. A faster camera might overload processing, and high-speed object detection is useless if decision-making modules lag behind. A true EtoE latency optimization requires analyzing individual component timing and their interaction within the entire system. Considerations include data flow efficiency, alignment of processing speeds throughout the pipeline, and the subsystem's impact on overall vehicle reaction time. A holistic HLS DSE approach enables designers to identify bottlenecks, uncover hidden optimization potential, and ultimately minimize EtoE reaction times, leading to more efficient autonomous driving systems.

This paper presents \textit{End-to-End Design Space Exploration (EtoE-DSE)}, a novel holistic system-level HLS DSE approach. EtoE-DSE addresses both \textit{variable timing constraints}, wherein different components in the system have different timing requirements, and EtoE latency constraints for the whole system, while efficiently identifying Pareto-optimal configurations in terms of energy and area for application-specific embedded systems. EtoE-DSE comprises a three-step process: an \textit{EtoE latency estimation model} to identify all required paths between any two endpoints within the targeted systems using an \textit{EtoE pathfinding (EPF)} algorithm; a \textit{frequency-based design space segmentation (FDSS)} process to segment and prune the design space into manageable subspaces based on variable timing constraints; and a \textit{latency-constrained segmentation optimization (LCSO)} algorithm to efficiently and accurately explore the system-level Pareto-optimal design solutions.

We evaluate the EtoE-DSE approach using a real-world use case of an autonomous driving subsystem (ADS) \cite{jo2015developmentautonomous}. The system's multi-component variable latency configuration resulted in a massive design space with 5.29e+128 solutions. We applied multiple different variable timing constraints and EtoE latency constraints to the ADS and implemented the systems on an FPGA for rigorous practical evaluation. On average, compared to the original design space, the FDSS process reduced the design space by 2.15e+55$\times$ without reducing the quality of results (QoR). The EtoE-DSE approach (EPF+FDSS+LCSO) rapidly, effectively, and accurately found the system-level Pareto-optimal configurations under EtoE constraints, improving the QoR by up to 89.26\% compared to the state-of-the-art in HLS DSE.

\begin{figure*}[th!]
\centering
\includegraphics[width=0.85\textwidth,keepaspectratio]{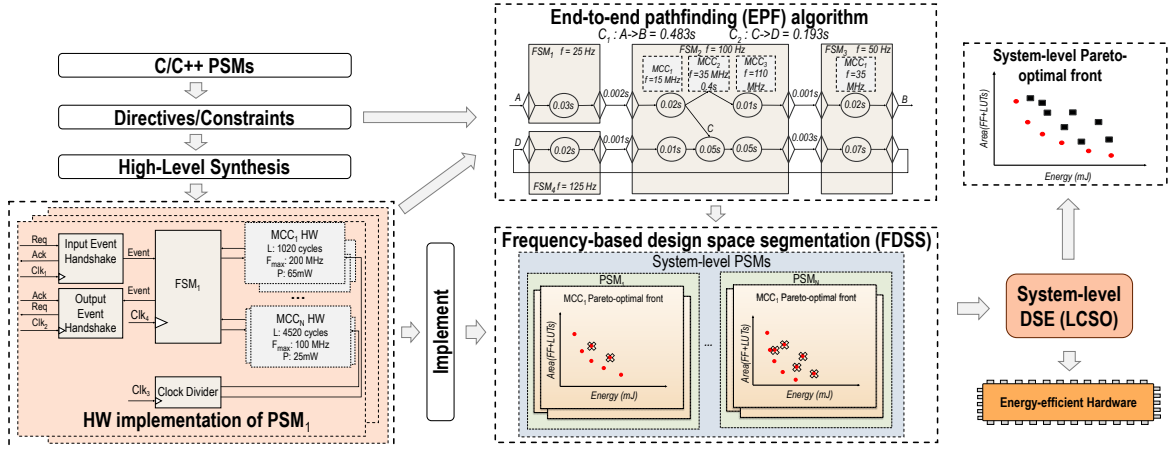}
\caption{The proposed \textit{End-to-End Design Space Exploration (EtoE-DSE)}  methodology to find system-level Pareto-optimal configurations while meeting the variable timing and end-to-end (EtoE) latency constraints.}
\label{fig:methodology}
\vspace{-10pt}
\end{figure*}

\section{Background and Related Work}
Most HLS research focuses only on the performance, power, and area, or a combination, instead of the timing requirements of embedded systems. HLS tools like Vitis HLS\cite{vivadoHLS2023}, SmartHLS\cite{SmartHLS2023}, Intel HLS\cite{IntelHLS2023}, and Bambu\cite{ferrandi2021bambu} use untimed C/C++ as input, producing hardware implementations with a targeted frequency. Some tools provide latency in terms of the number of cycles as one of the synthesis directives. For example, VivadoHLS allows designers to set up the number of cycles for function, loop, and output. However, designers still have limited control over the timing of the input source code. Liao et al.\cite{LIAOHHLS2022} proposed Hybrid-HLS and used a periodic state machine (PSM) model as an input structure to enable support for latency specifications directly in the input source code. However, an important limitation of this work is that it does not inherently support asynchronous systems. These systems may have multiple components potentially with different timing constraints, needing to communicate while still meeting overall system-level timing constraints.

On the other hand, many works\cite{LiaoDSE2023, GOSWAMI2023116, gao2021effective, SohrabizadehAutoDSE2021, YeAutoscaleDSE2023, Schafer_HLSDSE2020, Fer2008GA_HLS_DSE, wu2022ironman, gautier2022sherlock} have been proposed in High-Level Synthesis (HLS) Design Space Exploration (DSE) to efficiently and accurately find Pareto-optimal configurations. Conventional approaches often employ heuristic-based algorithms\cite{LiaoDSE2023, gao2021effective, Schafer_HLSDSE2020, Fer2008GA_HLS_DSE, SohrabizadehAutoDSE2021} to explore the design space, or machine learning-based techniques\cite{GOSWAMI2023116, wu2022ironman, gautier2022sherlock, YeAutoscaleDSE2023} to predict the objective variables (performance, power, and area) and speed up the DSE process. These strategies aim to uncover the optimal or Pareto-optimal configurations typically within single-component systems with single latency constraints. Unlike prior work, our approach focuses on providing support for variable timing (e.g., different periods or execution times) and end-to-end latency constraints \cite{mubeen2019supporting}. This enables the optimization of multi-component application-specific embedded systems where the different components might have different timing constraints. At the same time, our approach maintains the ability to efficiently and accurately find the Pareto-optimal system-level solutions.

\section{End-to end design space exploration (EtoE-DSE)}
Fig. \ref{fig:methodology} presents the proposed end-to-end design space exploration (EtoE-DSE) approach, a novel system-level HLS DSE method for precisely-timed multi-component application-specific embedded systems under EtoE latency constraints. Unlike existing techniques, this approach holistically estimates timing to meet variable timing and EtoE latency constraints while finding Pareto-optimal solutions. We first present the required system formalism to construct the desired precisely-timed, multi-component systems with \textit{Periodic State Machine} (PSM), \textit{Multi-cycle Computation} (MCC), and \textit{Event Handshake} components. We then present a novel timing estimation model that includes an \textit{EtoE pathfinding (EPF)} algorithm to find the correlated paths and sub-paths under different EtoE latency constraints and then estimate EtoE latencies for each path. Thereafter, we present a \textit{frequency-based design space segmentation (FDSS)} process to efficiently prune the original design space. Finally, we present a dedicated \textit{latency-constrained segmentation optimization (LCSO)} that uses an elite genetic algorithm to identify system-level Pareto-optimal solutions that satisfy the variable timing and EtoE latency constraints in terms of area and energy.

\begin{figure}[t]
\centering
\includegraphics[width=0.7\columnwidth,keepaspectratio]{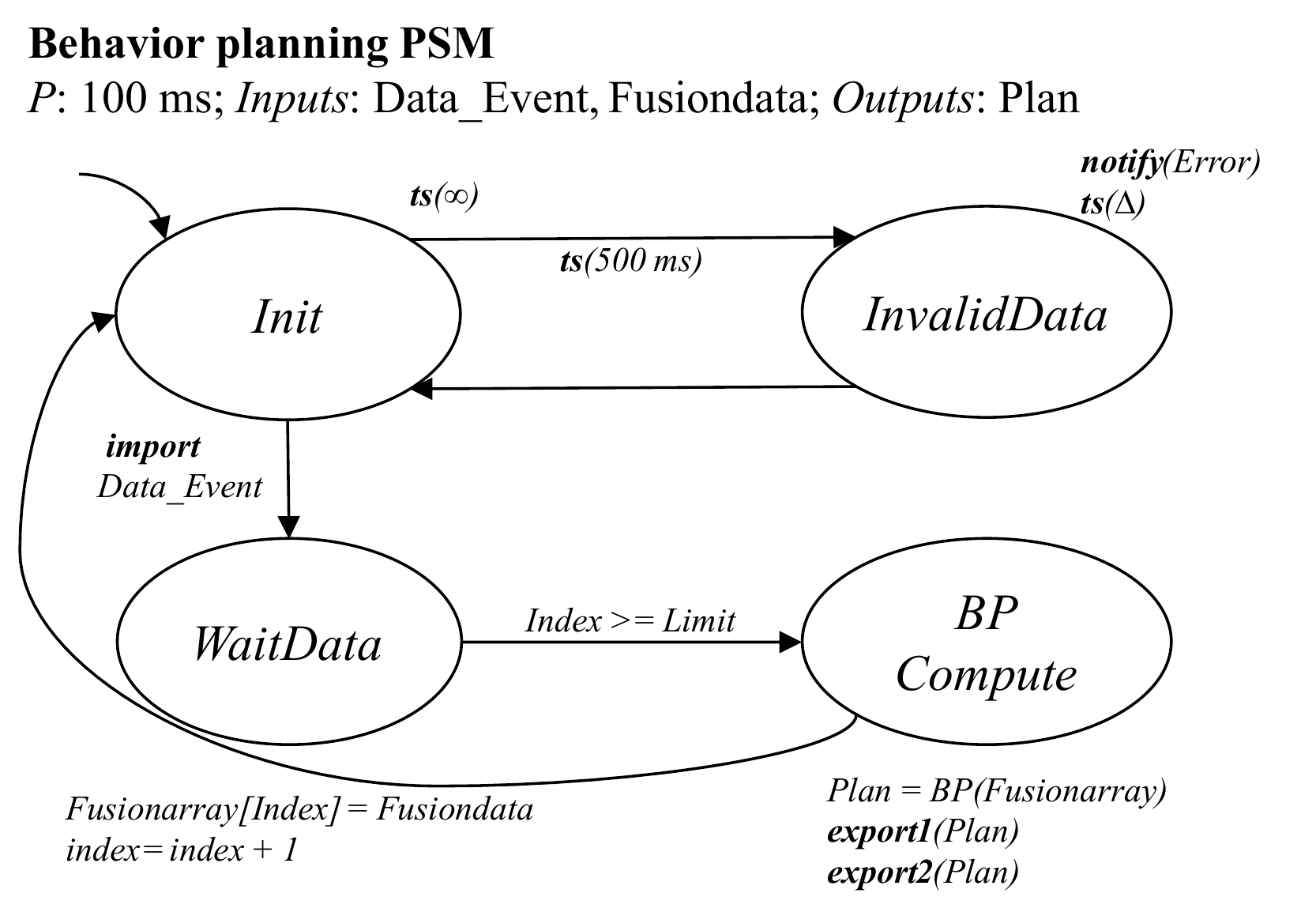}
\vspace{-5pt}
\caption{An example of a Periodic State Machine (PSM) model using a behavior planning component in the autonomous driving subsystem.}
\label{fig:BPPSM_highlevel}
\end{figure}

\begin{figure*}[t]
\centering
\includegraphics[width=0.85\textwidth,keepaspectratio]{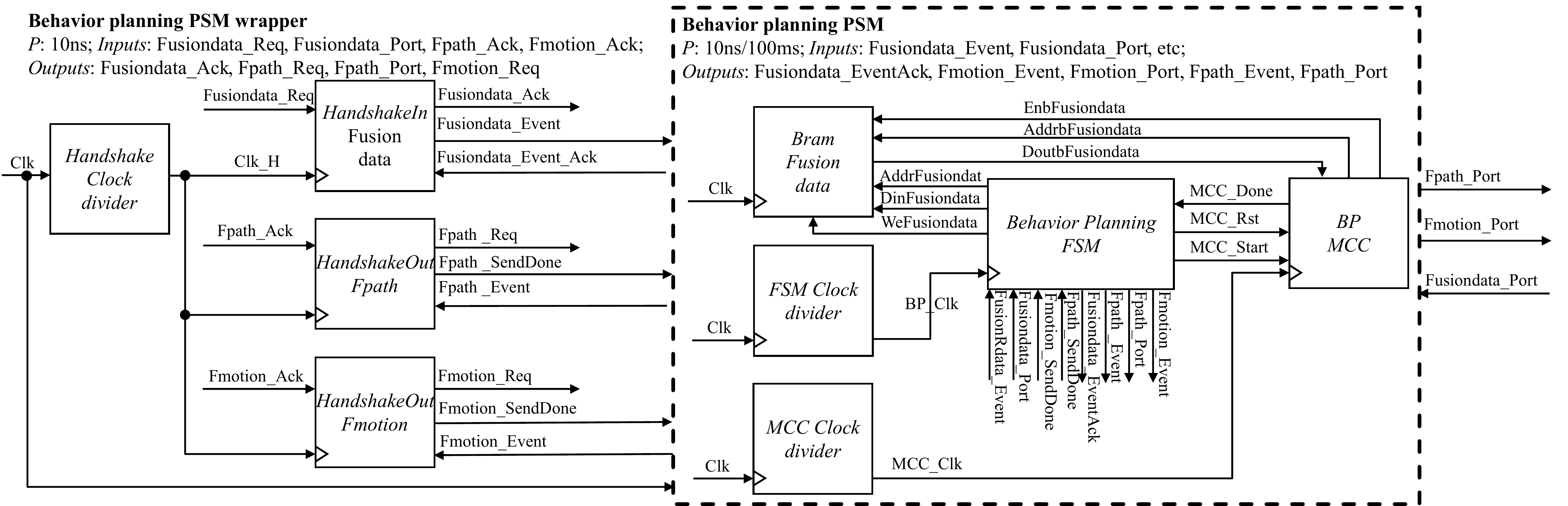}
\caption{A block diagram for the behavior planning PSM in an autonomous driving subsystem to illustrate the relationship between PSM, FSM, MCC, and both \textit{HandshakeIn} and \textit{HandshakeOut} components.}
\label{fig:BPPSM}
\end{figure*}

\subsection{System formalism}
EtoE-DSE focuses on multi-component, application-specific embedded systems, necessitating a precisely timed model formalism to construct such systems. This formalism enables designers to properly define the system structure, its components, and both component-level and system-level timing constraints. Here, we leverage \textit{Periodic State Machine} (PSM), \textit{Multi-Cycle Computation} (MCC), and \textit{Event Handshake} components as the main elements in the proposed model formalism.

\subsubsection{Periodic State Machine (PSM) and Multi-Cycle Computation (MCC)}
In EtoE-DSE, we use \textit{Periodic State Machines} (PSMs) \cite{kopetz07_PSM} as the component-level model formalism. Each system comprises one or more components, each represented as a PSM. PSMs are similar to \textit{Finite State Machines} (FSM), a conventional model for delineating computer systems' behavior. However, PSMs extend FSMs with features that enable the incorporation of time-triggered execution. This is dictated by specific elements such as a fixed period, clock constraints, a unified global time, and time-driven events. Each PSM comprises one or more multi-cycle computations (MCCs) that perform specific complex computations (e.g., median filter, matrix multiplication) that significantly influence the system's execution time. The PSM's fixed period is the first variable timing constraint that limits all MCCs in the PSM to be executed within this period. Different PSMs may have different periods, defined by the user to satisfy precise timing constraints, or depending on the timing requirements of their MCCs. As such, PSMs facilitate the high-level definition of intricate embedded systems that comply with precise timing constraints. Concurrently, they preserve the state-based characteristics intrinsic to FSMs, a feature that accommodates the Register Transfer Level (RTL) conversion of system specifications. Fig. \ref{fig:BPPSM_highlevel} shows an example PSM diagram for a behavior planning component in the autonomous driving subsystem (ADS). One MCC connects to the \textit{BP compute} state in the behavior planning PSM.

A PSM's specification can be expressed in a high-level language (e.g., C, C++), and HLS can be used to generate the corresponding RTL instantiation of the PSMs. This process forms FSMs integrated with custom datapaths. In each PSM, the MCCs are identified to create design alternatives through HLS. The design alternatives are functionally equivalent hardware implementations with different characteristics (e.g., different loop unrolling factors, maximum frequency) that yield tradeoffs in area, energy, and latency. The choice of MCC design alternatives significantly affects the efficiency of the system-level design. State-of-the-art HLS tools, like Xilinx Vitis, can generate different hardware implementations for MCCs based on a variety of directives/constraints, including critical path, maximum frequency, loop unrolling factor, etc. In our work, we assume a timing specification within the input code to specify variable timing and EtoE latency constraints. These constraints can be in the form of MCC execution latencies (e.g., the deadline for a computation or function), PSM periods, and system-level end-to-end (input-to-output and state-to-state) latency constraints, or timing resource constraints (e.g., number of frequency pins on the target FPGA). Tools like Vivado can then estimate performance metrics such as power consumption, critical path, and execution cycles for subsequent system-level design space exploration.

\begin{figure}[t]
\centering
\includegraphics[width=0.8\columnwidth,keepaspectratio]{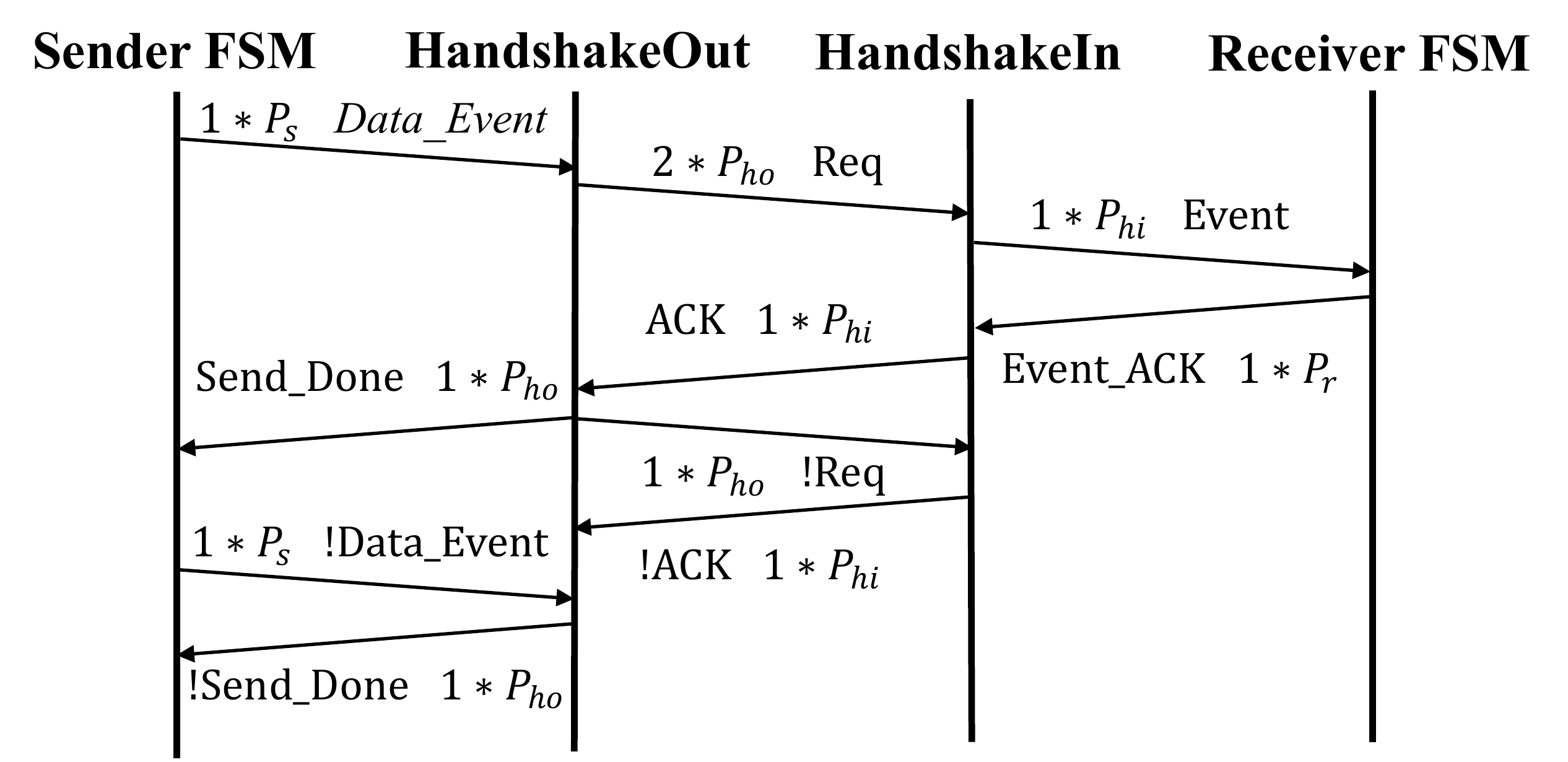}
% \vspace{-10pt}
\caption{A simplified illustration of an event handshaking protocol between sender PSM and receiver PSM.}
\label{fig:PSMhandshake}
\end{figure}

\subsubsection{Event Handshake}
The targeted application-specific embedded systems require communication and synchronization of data transfer events between the generated FSMs. We use a handshake-based technique \cite{van1992handshake}, which enables simplicity, reliability against data loss and design flexibility for different systems. Our approach features two handshake components: \textit{HandshakeIn} for receiving events and \textit{HandshakeOut} for sending events. Fig. \ref{fig:BPPSM} shows a block diagram for the behavior planning PSM in the ADS to illustrate the relationship between PSM, FSM, MCC, and both \textit{HandshakeIn} and \textit{HandshakeOut} components. The low-level components (MCCs, handshaking events, etc.) are extracted from the high-level PSM description of the component (Fig. \ref{fig:BPPSM_highlevel}) to provide the details for synthesis and RTL implementation. Fig. \ref{fig:PSMhandshake} illustrates the handshaking process. The receiver's \textit{HandshakeIn} receives the request from the sender's \textit{HandshakeOut} component and sends a signal to the receiver's FSM, then waits for a completion signal. Once the receiver's FSM finishes processing the event, the receiver's \textit{HandshakeIn} replies with an acknowledgment signal to the sender's \textit{HandshakeOut} component. The sender's FSM then proceeds with the next event after the completion signal is triggered from the sender's \textit{HandshakeOut}.

\begin{figure}[t]
\centering
\includegraphics[width=0.8\columnwidth,keepaspectratio]{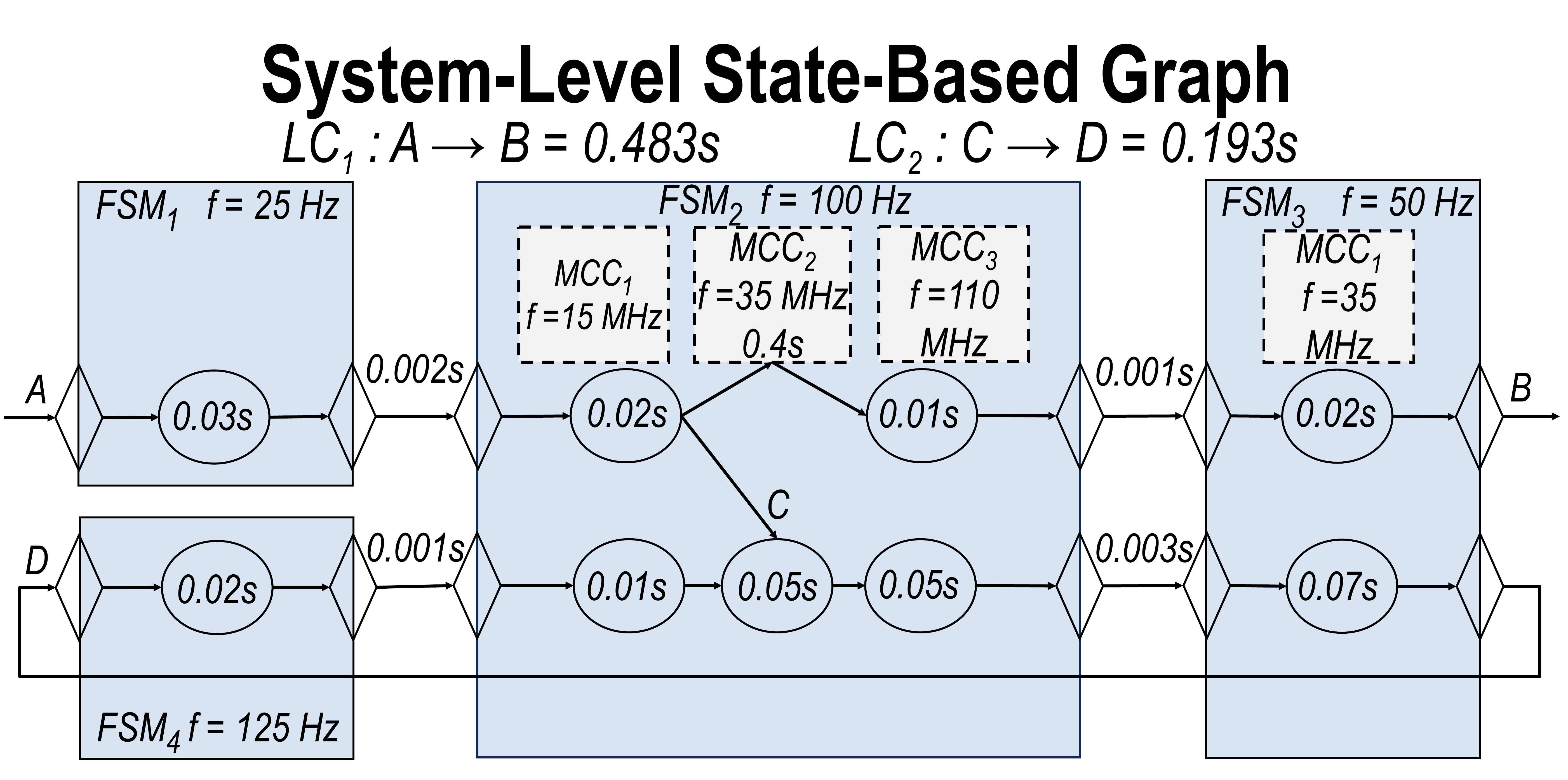}
\caption{An example of end-to-end (EtoE) latency and variable timing constrained system-level state-based graph. }
\label{fig:EtoE_graph}
\end{figure}

\subsection{End-to-end latency estimation model}
The second part of our EtoE-DSE approach estimates the EtoE latency for application-specific embedded systems. This estimated latency will be used during the subsequent system-level HLS DSE process. The EtoE latency estimation comprises three steps. The first step involves reading through the input PSMs, generated FSMs from HLS, MCCs, and potential handshake components (see Fig. \ref{fig:BPPSM}) to construct a system-level, state-based graph to represent the system. The second step explores the system-level graph to identify all EtoE paths and sub-paths. The final step of the EtoE latency estimation model calculates the latency for each path, taking into account the EtoE latency constraints.

\subsubsection{EtoE graph}
The system-level state-based graph---the \textit{EtoE graph}---employs a formalism similar to that of the PSM. However, to enable a high accuracy in latency estimation, we construct the graph with a holistic consideration of the FSMs generated from HLS tools for each PSM, the MCCs, and the handshake components. This enables us to precisely capture the interactions and dependencies between the components, and identify critical paths and sub-paths within the system that impact the latency most significantly. Fig. \ref{fig:EtoE_graph} presents an example graph where two EtoE latency constraints ($A \rightarrow B$ and $C \rightarrow D$) are applied to the system. Each FSM is associated with different PSMs, with the diamond shape denoting the handshake components. From endpoint A to B, the path passes through $FSM_1$, $FSM_2$, $MCC_2$ in $FSM_2$, and $FSM_3$. The sub-paths include any path within these EtoE latency constraints. For instance, in $FSM_1$, the sub-path from \textit{handshakeIn} to \textit{handshakeOut} has a latency constraint of $0.03s$. In $FSM_2$, the sub-path involving \textit{handshakeIn}, $MCC_2$, and \textit{handshakeOut} has a latency constraint of $0.02s + 0.4s + 0.01s$. The variable timing constraints applied to this path include the frequency of each FSM, MCC, and handshake component. From endpoint A to C, two paths are available. The first path leads from $FSM_1$ to $FSM_2$, while the second path goes through $FSM_1$, $FSM_2$, $FSM_3$, and $FSM_4$, passes D and then returns to $FSM_3$ at C. The sub-paths and alternative paths are taken into account in the path-finding algorithm.

\begin{algorithm}[t]
\DontPrintSemicolon
\scriptsize
\caption{EtoE pathfinding (EPF) algorithm}\label{alg:EtoE}
\KwIn{System-level graph $G$, EtoE latency constraints $LC_n$}
\KwOut{End-to-end paths and subpaths $Paths$}
\For{each FSM in FSMs in $G$}{
    \For{each State $S$ in FSM}{
        Set $S_{curr}$ to current state\;
        Call \textit{FindEtoEPaths} with ($S_{curr}$, \textit{tempPath}, \textit{Paths})\;
        Clear \textit{tempPath}\;
    }
}

\For{each path in $Paths$}{
    \For{each $LC$ in $LC_n$}{
        \If{path is not matched with all $LC$}{
            Delete path from $Paths$\;
        }
    }
}

\SetKwFunction{FMain}{$FindEtoEPaths$}
\SetKwProg{Fn}{function}{:}{}
\Fn{\FMain{$S_{curr}$, \textit{tempPath}, \textit{Paths}}}{
    Add $S_{curr}$ to \textit{tempPath}\;
    \eIf{$S_{curr}$ has no next states}{
        Create a new EtoE path $Path_{new}$ from \textit{tempPath}\;
        Store $Path_{new}$ in $Paths$\;
    }{
    \For{each $S_{next}$ of $S_{curr}$}{
        \eIf{$S_{next}$ is repeated in the \textit{tempPath}}{
            Create a new EtoE path $Path_{new}$ from \textit{tempPath}\;
        }
        {
            Call \textit{FindEtoEPaths} with ($S_{next}$, \textit{tempPath}, \textit{Paths})\;
        }
    }
    }
    Remove the last state from \textit{tempPath}\;
\KwRet \;
}
\end{algorithm}

\subsubsection{EtoE pathfinding (EPF) algorithm}
For more complex systems, like the autonomous driving subsystem considered herein, there may be multiple paths that can satisfy a single EtoE latency constraint while maintaining functional correctness. As such, the proposed EtoE-DSE approach employs a backtracking-based algorithm to discover alternative paths and sub-paths under each EtoE latency constraint. Although many pathfinding algorithms exist, such as A*, Dijkstra, and the maximum flow algorithm, the need for factoring EtoE latency constraints while identifying all paths and sub-paths necessitates the new \textit{EtoE pathfinding (EPF)} algorithm proposed herein. Rather than just finding the shortest or most efficient path, our EPF algorithm identifies the required paths and sub-paths that satisfy EtoE latency constraints for subsequent steps to estimate the latency. EPF is an essential element that enumerates the possible paths and does not affect the Quality of Results (QoR) of the EtoE-DSE approach.
 
Alg. \ref{alg:EtoE} presents pseudocode for the EPF algorithm. EPF's input is the system-level graph $G$ and EtoE latency constraints $LC_n$. It outputs the EtoE paths and sub-paths as a list of EtoE Path nodes, with each node containing the start state, end state, EtoE latency constraint, and paths between the two states. To identify the requisite paths, EPF invokes the recursive function \textit{FindEtoEPaths} for each state of the FSMs, ensuring that paths originating from a start state with no preceding state are considered (lines 1 - 7). For each state passed to \textit{FindEtoEPaths}, EPF identifies and stores paths that conclude at an end state with no subsequent state or a repeated state in the current $tempPath$. During each recursion, \textit{FindEtoEPaths} iterates through the current state $S_{curr}$’s next state $S_{next}$ and calls \textit{FindEtoEPaths} (lines 15 - 30). After finding all paths and sub-paths, EPF reviews the EtoE latency constraints to eliminate paths that do not meet the constraints and to remove redundant paths, ensuring only unique and relevant paths are retained (lines 8 - 14).

\subsubsection{EtoE latency estimation}
The last step of the EtoE latency estimation model estimates the worst-case latency for each path identified by EPF, considering variable timing constraints. For each path, the model initially identifies the assigned frequencies for each FSM, MCC, and handshake component. It then calculates the latency spent in each part by multiplying the number of states each FSM passes along the path, the execution cycles of the MCCs, and the communication cycles taken by each \textit{handshakeIn} and \textit{handshakeOut} component by their assigned frequencies. The final EtoE latency is the sum of all these individual latencies, and the EtoE latency for one path can be calculated as: 

\begin{equation}
\begin{split}
\small
& L_{path} = \sum_{i=1}^{n} L_{{FSM}_i} + \sum_{j=1}^{m} L_{{MCC}_j} + \sum_{k=1}^{p} L_{{handshake}_k}.
\end{split}
\end{equation}\label{eqa:EtoElatencycalculation}
\vspace{-5pt}

\noindent where $L$ represents the latency of each component. Exploring the worst-case cycles of handshake communication is needed for estimating the handshake latency. Given that the range of communication scenarios was tractable, we exhaustively explored the worst-case cycles by constructing real communication scenarios between the sender and receiver PSMs. We used the following equations to estimate the worst-case cycles:

Worst-case latency with respect to the receiver FSM:
\begin{equation}\label{eqa:receiverlatency_worst_Phi=Pho}
L_{r} \:<\:  ceiling\biggl(\biggl(P_{s} + 2 \times P_{ho} + P_{hi} + P_r\biggl) / P_{r}\biggl). 
\end{equation}

Worst-case latency with respect to the sender FSM:
\begin{equation}\label{eqa:senderlatency_worst_Phi=Pho}
L_{s} \:<\:  ceiling\biggl(\biggl(2 \times P_{s} + 4 \times P_{ho} + 2 \times P_{hi} + P_r\biggl) / P_{s}\biggl), 
\end{equation}

\noindent where $L_r$ and $L_s$ are the receivers' and senders' handshake latency. $P_{s}, P_{ho}, P_{hi}, P_r$ are the periods of the sender, handshakeOut, handshakeIn, and receiver, respectively. We empirically validated the estimation equations against an exhaustive exploration of the communications between handshaking components in the ADS use case. It is important to note that the end-to-end latency estimation is system-specific and alternative communication configurations can lead to latency variations. However, our work provides a solid foundation for implementing a latency estimation model in a different context.

\begin{algorithm}[t]
\DontPrintSemicolon
\scriptsize
\caption{Frequency-based design space segmentation (FDSS)}\label{alg:FDSS}
\KwIn{$ec$, $cp$, $w$, $f_{max}$, and $a$ for each MCC, $M$ and alternative, $M_{alter}$, in each PSM $P$, $N_{fpin}$, $T$, $N_{seg}$}
\KwOut{Segmented and pruned subspace in new PSM for each frequency segmentation in $list_{seg}$}
$E, f_s, s, \gets  calculateParameter(ec, f_{max}, cp, w)$ \;
$list_{set_f}, N_{comb}=\{(f_1, \ldots, f_{N_{fpin}}),\ldots,(f_1, \ldots, f_{N_{fpin}})\} \gets$ calculate a set of all possible frequency combinations $Set_f$ with total number of $N_{comb}$, each $set_f$ contains $N_{fpin}$ frequencies\; 
$N_{elem}= \frac{N_{comb}}{N_{seg} - 1} \gets$ Calculate the number of frequency comb elements for each segmentation\; 

\For{$i \gets 1$ \textbf{to} $N_{comb}$}{
     Combine $list_{set_f}$ ranging from $i \times N_{elem}$ to $(i + 1) \times N_{elem}$ in $seg_f$\; 
     Remove duplicate frequencies in each $seg_f$\;
     Store $set_f$ to $list_{seg_f}$\;
}
\For{$i \gets 1$ \textbf{to} $N_{seg} - 1$}{
    \For{each $M_{alter}$ in each $M$ in each $P$}{
        $f_{M_{alter}} \gets $ closest frequency in $list_{seg_f}i$\; 
        \eIf{no valid $M_{alter}$ in $M$}{
            $i \gets i + 1$, continue\;
        }{\If{all $M_{alter}$ is valid}{
            $E_{M_{alter}} \gets f_{M_{alter}} \times s$\;
            sort all $M_{alter}$ in each $M$ based on Pareto-optimal $E_{M_{alter}}$ and $a$\; 
        }}
    }
    \If{all $M_{alter}$ in each $M$ in each $P$ sorted}{ 
        output new PSM\;
    }
    $i \gets i + 1$\;
}
\end{algorithm}

\subsection{Frequency-based design space segmentation (FDSS)}\label{sec:FDSSalgorithm}
The EtoE latency estimation model calculates the latency for each required EtoE path, providing a solution to verify the validity of design points during the HLS DSE process. Following the estimation, the EtoE-DSE approach undertakes a frequency-based design space segmentation (FDSS) process. Our primary motivation for choosing frequency-based pruning as the initial step, rather than alternative methods like resource-based pruning, is the impact of frequency on EtoE latency. A change in frequency can significantly affect the latency of each component within a path. The segmentation's objective is twofold: 1) FDSS divides the original design space into subspaces by segmenting the frequency combinations into a manageable number for parallelization based on the number of available resources (e.g., cores); and 2) FDSS preemptively prunes or eliminates sub-optimal solutions in term of energy and area. This is achieved by assigning frequencies from each segment within the subspaces to MCC alternatives in each PSM. Such segmentation and pruning efforts make the DSE process more tractable and less time-consuming. By dividing the design space into subspaces, the system-level DSE algorithm can more quickly identify solutions that enhance the target objective functions, such as area and energy consumption. Specifically, for timing-constrained multi-component embedded systems, the proposed FDSS algorithm seeks to segment the design space and prune solutions that fail to meet the variable timing constraints associated with frequency selection for MCCs. 

Alg. \ref{alg:FDSS} presents the pseudocode for the FDSS algorithm. The algorithm's inputs are the execution cycles $ec$, maximum frequency $f_{max}$, critical path $cp$, power $w$, and area $a$ for each MCC alternative, denoted as $MCC_{alter}$. Additionally, three critical input constraints for FDSS include the number of allowed clock frequencies $N_{fpin}$ for the embedded system, constrained by the target hardware---we focus on FPGAs---the assigned period $T$ for each PSM based on the variable timing constraint, and the segmentation number $N_{seg}$, which depends on the number of threads that EtoE-DSE may use for parallelization. FDSS outputs the segmented and pruned subspace for the input system. To segment and prune the original design space, FDSS first calculates the estimated energy $E$, a scaled frequency $f_s$, and a scaling factor $s$. The scaled frequency represents the minimum frequency at which an MCC alternative can operate under $T$, and the scaling factor, calculated as $\frac{w}{f_{max} \times T}$, is used to estimate the new energy when applying a new frequency to an MCC alternative. Subsequently, with $N_{fpin}$, the maximum of the minimum frequency, and the maximum frequency for each MCC in the PSMs, FDSS explores a list of possible frequency combination sets $list_{set_f}$ (lines 1 - 2) by iterating through them, where each $set_f$ contains $N_{fpin}$ frequencies. Next, FDSS calculates the number of frequency combination sets $N_{elem}$ that should be segmented and add them together based on the segmentation number $N_{seg}$ (line 3). With all parameters now prepared, FDSS is ready for operation.

Before FDSS can create subspaces in the format of a PSM, it must first segment the frequency combination sets. Each segment includes $N_{elem}$ frequency combination sets, totaling $N_{seg} - 1$ segments, which are stored in $list_{set_f}$ (lines 4 - 8). The last segmentation is reserved for the original design space and the original frequency set to prevent data loss under tight EtoE latency constraints; this will be explained in detail in the next section. FDSS then proceeds to create subspaces. In each iteration $i$, FDSS reviews every MCC alternative, locates the closest frequency in $list_{set_f}i$ to each MCC alternative's minimum frequency (lines 9 - 11), and assigns the new frequency $f_{M_{alter}}$ to the MCC. If an MCC alternative cannot match a valid closest frequency, FDSS moves to the next segment $list_{set_f}i$ (lines 12 - 13). Conversely, if $f_{M_{alter}}$ is valid for all MCC alternatives, FDSS calculates the current energy $E_{M_{alter}}$ for each by multiplying $f_{M_{alter}}$ by the scaling factor $s$. Utilizing $E_{M_{alter}}$ and the area $a$, FDSS prunes MCC alternatives to identify Pareto-optimal alternatives for each MCC (lines 14 - 19). Finally, FDSS outputs a set of segmented and pruned subspaces in terms of a new PSM corresponding to each valid frequency segment (lines 21 - 23). After the FDSS process, the latency-constrained segmentation optimization (LCSO) is then applied to identify the system-level Pareto-optimal solutions.

\subsection{Latency constrained segmentation optimization (LCSO)}\label{sec:LCSO}
Although the FDSS algorithm significantly reduces the original design space by eliminating less optimal MCC alternatives in each frequency segment, the design space at the system level remains extremely large. Employing the PSM formalism and the selected frequency segment from an allowable range of frequencies ensures that the potential solutions from the segmented and pruned subspaces have been determined to meet the variable timing constraints for PSMs (under period $T$). However, we are still faced with a multi-objective optimization problem, considering the potentially conflicting objectives of energy and area against the backdrop of EtoE latency constraints and variable timing constraints on frequency selection for MCC, FSM, and handshake components. To tackle this challenge, we introduce a \textit{Latency-Constrained Segmentation Optimization} (LCSO) process that leverages an elite genetic algorithm (GA) within the EtoE-DSE approach.

GAs are frequently used in addressing multi-objective High-Level Synthesis (HLS) Design Space Exploration (DSE) challenges \cite{gao2021effective,pilato2010speeding_GA,Fer2008GA_HLS_DSE}. Typically, a GA operates through an iterative process involving a population of potential solutions. These solutions undergo iterative refinement toward optimal outcomes. This refinement process evaluates the fitness of each solution, selects the fittest solutions through stochastic methods, and employs random mutation and crossover techniques to generate subsequent generations \cite{Deb2002elitGA}. The versatility of GAs aligns well with our objectives, particularly due to the complex, multi-modal PSM search space and the need to balance competing objectives within this space. Additionally, the population-based strategy of GAs inherently provides greater design versatility, enabling the generation of diverse designs that meet varying, user-defined constraints (variable timing constraints and EtoE latency constraints).

The GA used in the LCSO process is distinguished by three important features aimed at improving the GA's overall effectiveness. Firstly, our approach incorporates \textit{dynamic Pareto-optimal elitism}. Unlike conventional methods that maintain a static size for the elite population, our strategy ensures the preservation and carryover of Pareto-optimal solutions from one generation to the next, ensuring the solution quality across generations. Secondly, we adopt a genetic representation in which the input structure contains important information uniquely suited for the target systems' complexity. This genetic representation encapsulates the selected MCC alternative for each MCC, the handshake component, and the FSM with an associated frequency. This frequency is randomly selected from the frequency segment, as described in Section \ref{sec:FDSSalgorithm}, to serve as a gene. The MCC alternatives, handshake components, and FSM in each PSM form the first part of a chromosome $C_r$, and the selected frequencies from each frequency segment forms the second part of $C_r$. Third, we leverage the inherent parallelism of GAs by separating the segmented and pruned subspace into different threads or a single thread with a streamlined GA configuration. This streamlined configuration comprises the number of populations, generations, and segments, and other hyperparameters are determined and remain unchanged following a careful evaluation, which achieves an improved QoR. The details of the GA configuration are discussed in Section \ref{sec:experiments}. When FDSS segments the design space and generates new PSMs based on each frequency segmentation, each new PSM can be assigned to a different thread during the DSE process, thereby reducing the execution time for exploring large design spaces.

\begin{algorithm}[t]
\scriptsize
\DontPrintSemicolon
\caption{Latency constrained segmentation optimization (LCSO)}\label{alg:LCSO}
\KwIn{$Paths$, $T$, $p_s$, $p_c$, $p_m$, $k$, and $seg_f$, $E$, $a$ from segmented and pruned PSMs $PSM_{seg}$}
\KwOut{Pareto-optimal system-level solutions}
\For{\textbf{each} valid $seg_f$ and $PSM_{seg}$}{
    $P_k, flag \gets initPop(k, seg_f, PSM_{seg}, Paths)$ initial a population of $k$ randomly-generated individuals that satisfies $LC_n$ in each $Paths$\; 
    \If{$!flag$}{$continue$\;}
    \While{!terminate condition}{
    new $P_k \gets ParetoElite(P_k,eliteP_k)$\;
    %fitness score 
    $S_k \gets fitness(P_k)$\;
    %select population 
    $PS_k \gets select(P_k,S_k,p_s)$\; 
    %crossover population 
    $PC_k \gets crossover(PS_k,p_c)$\; 
    %mutate population 
    $PM_k \gets mutate(PC_k,p_m)$\;
    }
    %save Pareto-optimal configurations across each pruned PSM 
    $savePareto(eliteP_k)$\; 
}
\SetKwFunction{FSub}{$InitPop$}
\SetKwProg{Fn}{function}{:}{}
\Fn{\FSub{$k, seg_f, PSM_{seg}, Paths$}}{
    \While{$size(P_k) != k$}{
        $C_r \gets$ randomly generate one $C_r$\;
        \If{$C_r$ is valid for all $Paths$}{
            Insert $C_r$ to $P_k$\;
        }
        \If{$T$ is reached}{
            \KwRet $False$\;
        }
    }
\KwRet $P_k$\;
}

\SetKwFunction{FMain}{$ParetoElite$}
\SetKwProg{Fn}{function}{:}{}
\Fn{\FMain{$P_k,eliteP_k$}}{
$P_k \gets$ Insert old $eliteP_k$ into $P_k$\;
new $eliteP_k \gets$ Pareto-optimal population in $E$ and $a$ in $P_k$\;
\KwRet $P_k$\;
}
\end{algorithm}

Alg. \ref{alg:LCSO} depicts the pseudocode of the LCSO algorithm. LCSO takes as input the required EtoE latency constrainted paths $Paths$, select rate $p_s$, crossover rate $p_c$, mutation rate $p_m$, population size $k$, and frequency segment $seg_f$, energy $E$, area $a$ from MCC alternatives $M_{alter}$ in each the segmented and pruned PSM $PSM_{seg}$. The algorithm outputs the Pareto-optimal system-level configurations. LCSO starts by iterating through each valid $seg_f$ and its corresponding $PSM_{seg}$ (line 1). In each $PSM_{seg}$, LCSO calls \textit{InitPop} function to generate the initial population (line 2). \textit{InitPop} randomly generates an initial population $P_k$ with size $k$ that satisfies the required EtoE $Paths$. If the EtoE latency constraints are too tight, the initialization process may never find valid design solutions. LCSO sets a threshold $T$ to stop the \textit{InitPop} and abandons current frequency segment if no valid solutions are found (line 12 - 19). This $T$ is determined by the two times the maximum number of chromosomes in all valid segment subspace used to fulfill the initial population. Next, LCSO iterates until a terminate condition is met. The terminate condition can be a predefined number of generations, which we use in this work, or a threshold of the quality of results (e.g., the distance from a reference set). In each generation, for the $ParetoElite$ function, LCSO goes through the current population $P_k$ and the previous generation’s elite population $eliteP_k$  (initially an empty set) to find and save the current Pareto-optimal $C_r$ to the new $eliteP_k$ set. Unlike typical elite functions with a fixed elite member size \cite{paliwal2019reinforced_elite}, LCSO stores every Pareto-optimal $C_r$. $ParetoElite$ outputs the new population $P_k$. If the new $P_k$ exceeds size k, any $C_r$ with the least fit score is discarded until the population size constraint is satisfied (lines 26 - 29). To calculate the fitness of the population, we use a custom fitness function defined in Equation \ref{eqa:fitness}, comprising the weighted sum of energy and area (line 8):

\begin{equation}
\label{eqa:fitness}
\begin{split}
& -1 \times (  A \log(\max(1 - \frac{E_i}{E_{max}}, 1 \times 10^{-5})) 
\\& + B \log(\max(1 - \frac{a_i}{a_{max}}, 1 \times 10^{-5}))),
\end{split}
\end{equation}
\vspace{-5pt}

\noindent where $A$ is the weight for energy $E$, $B$ is the weight for area $a$.

The rest of LCSO contains selection, crossover, and mutation (lines 9 - 11). LCSO uses the roulette-wheel method for selection \cite{shukla2015roulette_wheel}; for crossover, LCSO generates children $C_r$ until the population size equals $k$; and mutation randomly selects and changes the selected MCC alternative, and the frequency of MCC alternative, handshake component, and FSM. Finally, after the terminate condition is met, the $savePareto$ function saves the last $eliteP_k$ from each $P$ (line 13). $T$, $P_s$, $p_c$, $p_m$, and $k$ are hyperparameters that can be tuned to improve the search process for the target design space. In our experiments, we set $T = 10k$, $p_s = 0.5$, $p_c = 0.7$, and $P_m = 0.5$ after conducting multiple tests to identify the best Pareto-optimal front, while keeping other hyperparameters unchanged.

Although LCSO employs the threshold $T$ to terminate exploration in any segmented subspace where no valid population is found, EtoE-DSE must consider one worst-case scenario. This scenario arises when, under more stringent EtoE latency constraints, LCSO may fail to find any valid solutions within every segmented and pruned subspace generated by FDSS, even though the original design space contains valid solutions. To address this issue, EtoE-DSE substitutes any invalid segmented subspace with the original design space, thereby continuing the exploration for a system-level Pareto-optimal front.

\begin{figure}[t]
\centering
\includegraphics[width=0.8\columnwidth,keepaspectratio]{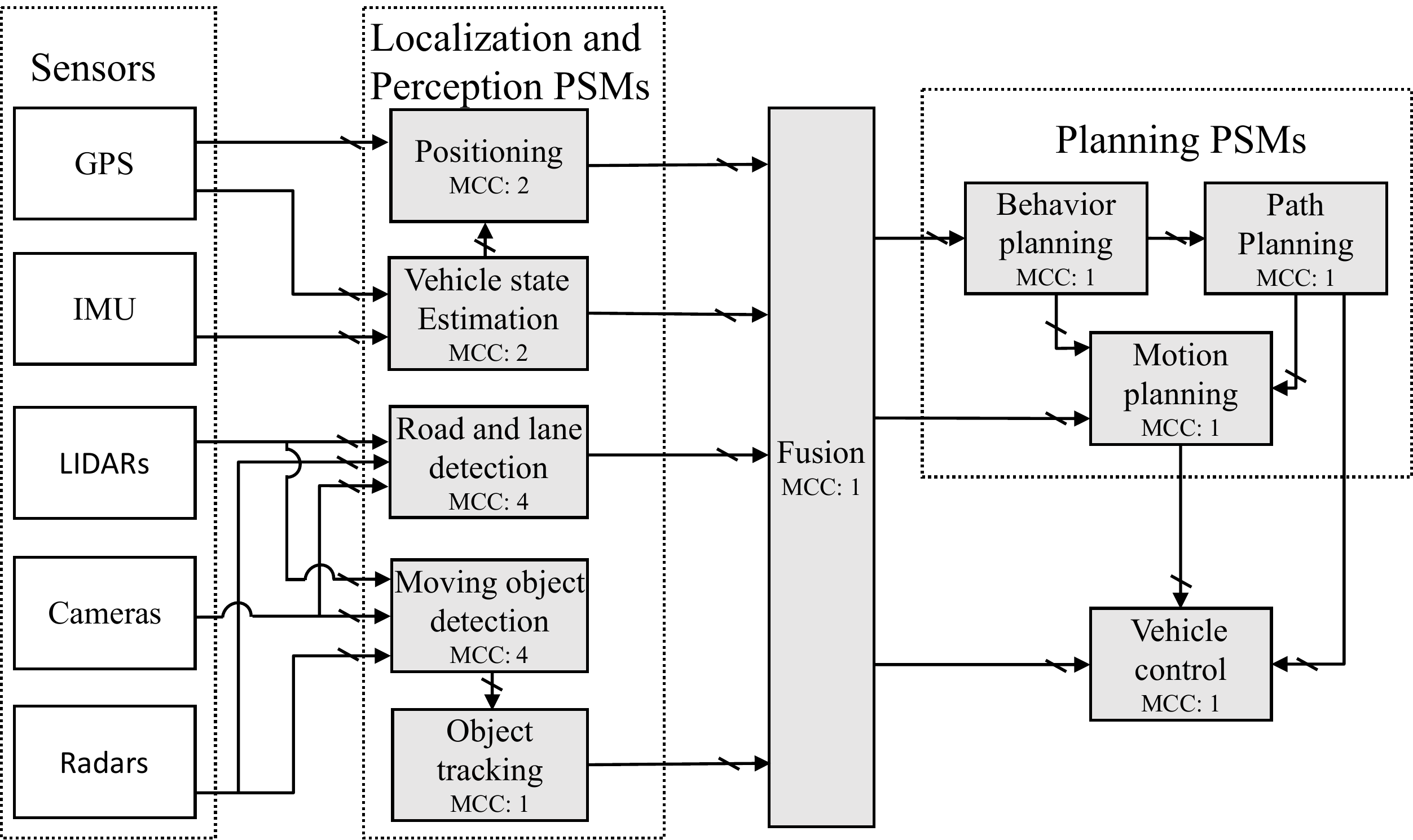}
% \vspace{-10pt}
\caption{Block diagram of the autonomous driving subsystem (ADS). Shaded blocks are the components implemented as PSMs in our experiments. Where each arrow represents a handshake-based event}
\label{fig:autonomousdriving}
%\vspace{-10pt}
\end{figure}

\section{Experiments} \label{sec:experiments}
To evaluate EtoE-DSE, we utilized a real-world use case of a complex autonomous driving subsystem (ADS) \cite{jo2015developmentautonomous}. Fig. \ref{fig:autonomousdriving} presents the ADS system-level block diagram, which comprises five main parts: \textit{sensors}, \textit{localization and perception}, \textit{method fusion}, \textit{planning}, and \textit{vehicle control}. Apart from the sensors, the remaining parts are implemented as PSMs. Each PSM generates an FSM and consists of 1 to 4 MCCs, where each MCC may have multiple alternative hardware implementations (accelerators) based on the input directives/constraints using HLS tools. In total, the ADS includes 10 PSMs, 18 MCCs, and 490 MCC alternatives. To delineate the original design space for ADS, specific constraints were established before the experiment. Initially, a frequency pin constraint of 4 was applied to the FPGA Artix-7 board with a part number XC7A100T. The minimum and maximum scaled frequencies for MCC alternatives, allowing for operation at speeds necessary under their respective PSM's period, are set at 2 and 108 MHz, respectively, while the original frequencies range from 90 to 240 MHz. The scaled frequency ensures MCC alternatives meet the first variable timing constraint, the period assigned for each PSM. For fine-grained frequency exploration, we selected a frequency interval of 5 MHz. This interval, along with the frequency range for MCC alternatives and the frequency pin constraint, results in 1329 frequency combinations available for each system-level configuration. 

Finally, considering the number of FSMs, handshake components, and possible frequencies, the original design space for ADS is calculated to be 5.291e+128. Such an astronomical design space necessitates an efficient DSE approach. To put this in perspective, even a smaller system like a three-component wearable pregnancy monitor has a design space of approximately 1e+10, when the different component-level and system-level design alternatives are considered. As the number of components, MCC alternatives, and frequency options increase, the design space grows exponentially. This vast design space underscores the critical need for an efficient methodology to identify Pareto-optimal solutions that can address the scalability challenges of complex systems like the ADS design.

The EtoE-DSE approach is implemented in C++ and run on an Intel Xeon CPU, using up to 48 threads. Initially, we evaluate the EtoE pathfinding algorithm (EPF) for the ADS system. We selected two sets of EtoE points: The first set begins at the input of the GPS PSM and ends at the output of the Control PSM; the second set starts at the input of the LIDAR PSM and ends at the output of the Fusion PSM. Subsequently, we evaluate the design space reduction achieved by Frequency-based Design Space Segmentation (FDSS). Finally, we evaluate the Quality of Results (QoR) for the Pareto-optimal solutions under EtoE latency constraints using EtoE-DSE, compared to the state-of-the-art represented by a Genetic Algorithm (GA), Simulated Annealing (SA), and Ant Colony Optimization (ACO) presented in \cite{gao2021effective}. Due to the nature of the problem resulting in a complex data structure for each solution, the ACO algorithm was unable to converge to satisfactory solutions, and was significantly outperformed by all other considered methods. Therefore, for a fair comparison, we only present the results of the GA and SA algorithms as proxies for prior work. We note that no prior work exists, to the best of our knowledge, that holistically considers the EtoE latency constraint of HLS DSE in multi-component systems as we do herein. Thus, we applied EPF to prior work and compared it with EtoE-DSE. 

For a robust evaluation, we created three variants of EtoE-DSE: 1) EtoE-DSE without FDSS (EtoE-DSE$_{unsegmented}$), to evaluate any performance loss from prior segmentation and pruning processes; 2) EtoE-DSE with serial LCSO running on a single core (EtoE-DSE$_{serial}$); and 3) EtoE-DSE with parallelized LCSO (EtoE-DSE$_{parallel}$). To further evaluate EtoE-DSE under varying EtoE latency constraints for the ADS system, we selected four constraints: 50 ms, 100 ms, 150 ms, and 200 ms for the 15 paths in the two EtoE pairs. Another crucial setup adjustment was made to the fitness function in Equation \ref{eqa:fitness}, where the weight $A = 2$ and $B = 1$ to emphasize energy over area, considering the significant role of estimated energy calculations in EtoE latency. Hence, prioritizing energy significantly impacts and modifies EtoE latency.

\begin{table}[t]
    \caption{Hyperparameter setting for the GA and SA algorithms (prior work), EtoE-DSE$_{unsegmented}$, EtoE-DSE$_{serial}$, and EtoE-DSE$_{parallel}$.}
    \label{tab:hyperparameter}
    \scriptsize
    \centering
    \begin{tabular}{cccc}
    \hline
    \textbf{Task} & \textbf{Population} & \textbf{Generation}  & \textbf{Segment}\\ 
    \hline
    \textit{Prior work (GA)} & 1000 & 1000 & 0 \\  
    \hline
    \textit{Prior work (SA)} & Temperature diff 850 & 2000 & blue{0} \\  
    \hline
    \textit{EtoE-DSE$_{unsegmented}$} & 1000 & 1000 & 0\\ 
    \hline
    \multirow{3}{*}{\textit{EtoE-DSE$_{serial}$}} & 230 & 100 & 44\\ 
    
    & 230 & 200 & 22\\ 
    
    & 230 & 400 & 11\\ 
    \hline
    \multirow{3}{*}{\textit{EtoE-DSE$_{parallel}$}} & 230 & 400 & 44\\ 
    
    & 230 & 800 & 44\\
    
    & 230 & 1600 & 44\\ 
    \hline
    \end{tabular}
    %\vspace{-10pt}
\end{table}

Table. \ref{tab:hyperparameter} depicts the hyperparameter settings for the streamlined GA configuration in LCSO, as mentioned in Section \ref{sec:LCSO}. The streamlined GA configuration is described as: $<$number of population ($k$)$>$-$<$number of generations ($G$)$>$-$<$number of segments ($N_{seg}$)$>$. Other hyperparameters, such as the selection rate, remain unchanged across prior work and EtoE-DSE. We selected three configurations for EtoE-DSE$_{serial}$ and EtoE-DSE$_{parallel}$. Ideally, the total number of chromosome changes across the entire algorithm should equal $k \times G \times N_{seg}$. For prior work and EtoE-DSE$_{unsegmented}$, we employed a base configuration of $1k-1k-0$, resulting in a total chromosome change of 1 million. We controlled this number in the experiment for EtoE-DSE$_{serial}$ to establish an equal environment and evaluate the extent of improvement achievable by EtoE-DSE. In contrast, the configuration for EtoE-DSE$_{parallel}$ differs due to the parallelization process, which supports more significant potential improvements in a shorter time than EtoE-DSE$_{serial}$.

The last critical factor involves establishing a suitable measure for the QoR of EtoE-DSE variants and prior work. The QoR is a score comparing the estimated Pareto-optimal front to the optimal solution front. However, due to the ADS system-level design space being extremely large, conducting an exhaustive search (ES) was prohibitive. Consequently, it is necessary to identify the closest reference set to the optimal front. To this end, we conducted multiple different runs for the three variants and prior work, merging all results to identify the reference Pareto-optimal front ($P_{ref}$) for each specific EtoE latency constraint. For each EtoE latency constraint, we created a different reference set because reference sets with softer constraints may not have valid solutions for more stringent constraints. The QoR is then calculated by comparing each estimated Pareto-optimal front with $P_{ref}$. Prior work used the \textit{average distance to the reference set} (ADRS) method \cite{Schafer_HLSDSE2020} for comparing the estimated Pareto-optimal front with $P_{ref}$. However, the ADRS metric has a drawback: if one of the objective variables consistently exhibits a larger percentage difference, the other objective variable will not be considered in the evaluation. To address this drawback, we modified the ADRS to create a new metric, the \textit{average Euclidean distance to the reference set} (AEDRS). This modification involves calculating the Euclidean distance of the percentage difference for both energy and area rather than solely accounting for the largest percentage difference for one objective variable. The AEDRS is given by the following equation:

\begin{equation}
% \small
AEDRS(\Phi, \Omega) = \frac{1}{|\Omega|} \sum_{\omega \in \Omega}{min}_{\phi \in \Phi}R^2(\phi,\omega),
\end{equation}\label{eqa:AEDRS}

with $R^2(\phi,\omega)$:

\begin{equation}
\small
R^2(\phi = (E_\phi, a_\phi), \omega = (E_\omega, a_\omega)) = \sqrt{(\frac{E_\omega - E_\phi}{E_\phi})^2 + (\frac{a_\omega - a_\phi}{a_\phi})^2}.
\end{equation}\label{eqa:Euclidean}

Here, $\Phi$ represents the reference set, and $\Omega$ denotes the estimated Pareto-optimal front. The Euclidean distance takes into account the differences in both energy ($E$) and area ($a$). A lower AEDRS score indicates that the estimated Pareto-optimal front is similar to the reference front, whereas a higher AEDRS score suggests that the estimated Pareto-optimal front differs more significantly from the reference front.

\begin{table}[b]
\vspace{2pt}
\scriptsize
\caption{ADS's design space sizes after applying FDSS algorithm with different segment threshold. The original design space is 5.291e+128}
% \vspace{-10pt}
\begin{center}
\begin{tabular}{ccc}
\toprule
% \textbf{Segment}&\multicolumn{2}{c}{\textbf{Design space}}\\
% \cline{2-3} 
\textbf{Segment} & \textbf{\textit{After FDSS}}& \textbf{\textit{Improvement}} \\
\midrule
44 & 2.469e+73 & 2.142e+55x \\

22 & 2.456e+73 & 2.154e+55x \\

11 & 2.443e+73 & 2.165e+55x \\

Geo. mean & 2.45629e+73 & 2.15389e+55x \\
\bottomrule
\end{tabular}
\label{tab:FDSSdesignspace}
\end{center}
\end{table}

\begin{figure}[t]
\centering
\includegraphics[width=1\columnwidth,keepaspectratio]{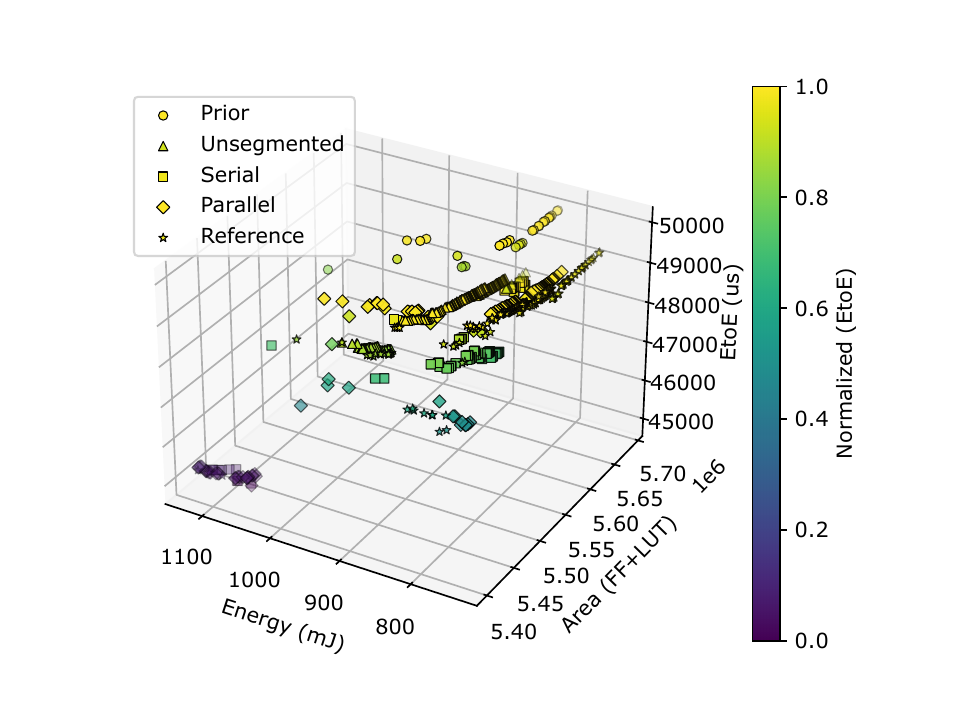}
\vspace{-10pt}
\caption{Visualized 3D comparison of the system-level Pareto-optimal fronts between reference set, and estimated front from prior work (GA), EtoE-DSE$_{unsegmented}$, EtoE-DSE$_{serial}$, and EtoE-DSE$_{parallel}$ under 50ms end-to-end (EtoE) latency constraint. Different level of color in color bar represents the EtoE latency for one of the paths between (GPSsensor to Control) PSMs}
\label{fig:3D5z_50ms}
\end{figure}

\section{Results}
\subsection{Path finding using EPF}
In total, EPF identified 8099 paths and sub-paths within the ADS. For example, for EtoE points from the GPS PSM input to the Control PSM output, EPF discovered 13 paths, and 2 paths for the EtoE points from the LIDAR PSM input to the Fusion PSM output. Note that at this stage, EtoE latency constraints are specified for these two sets of EtoE points but are not computed, as the frequencies for FSMs, MCCs, and handshake components will not be chosen until the LCSO algorithm identifies possible system-level configurations.

\subsection{Frequency segmentation and subspace pruning using FDSS}
Table \ref{tab:FDSSdesignspace} illustrates the design space reduction achieved by applying FDSS to the ADS system. We utilized three values of $N_{seg}$: 44, 22, and 11, to segment the 1329 frequency combinations. Subsequently, FDSS reduced the ADS design space from 5.29e+128 to 2.47e+73, 2.46e+73, and 2.44e+73 for the 44, 22, and 11 frequency segments, respectively. In comparison to the original design space, FDSS achieves an average reduction of approximately 2.15E+55 times. A key observation from the subspace analysis is that, regardless of the $N_{seg}$ value, the subspace size remains at a similar order of magnitude (e+73). This is primarily because, after segmenting by frequency and pruning the MCC alternatives to eliminate sub-optimal choices based on estimated energy and area, one of the frequency segments invariably contains the most frequencies. The pruned subspace of this segment attains the largest magnitude (e+73), and the addition of other subspaces does not alter this magnitude. Although FDSS significantly reduces the design space, its true merit lies in narrowing down the design space without discarding potential high-quality solutions from the search space. Therefore, in evaluating the overall EtoE-DSE approach, we also evaluate the QoR achieved with and without employing the FDSS algorithm, in relation to the system-level Pareto-optimal configurations.

\begin{figure}[t]
\centering
\includegraphics[width=0.9\columnwidth,keepaspectratio]{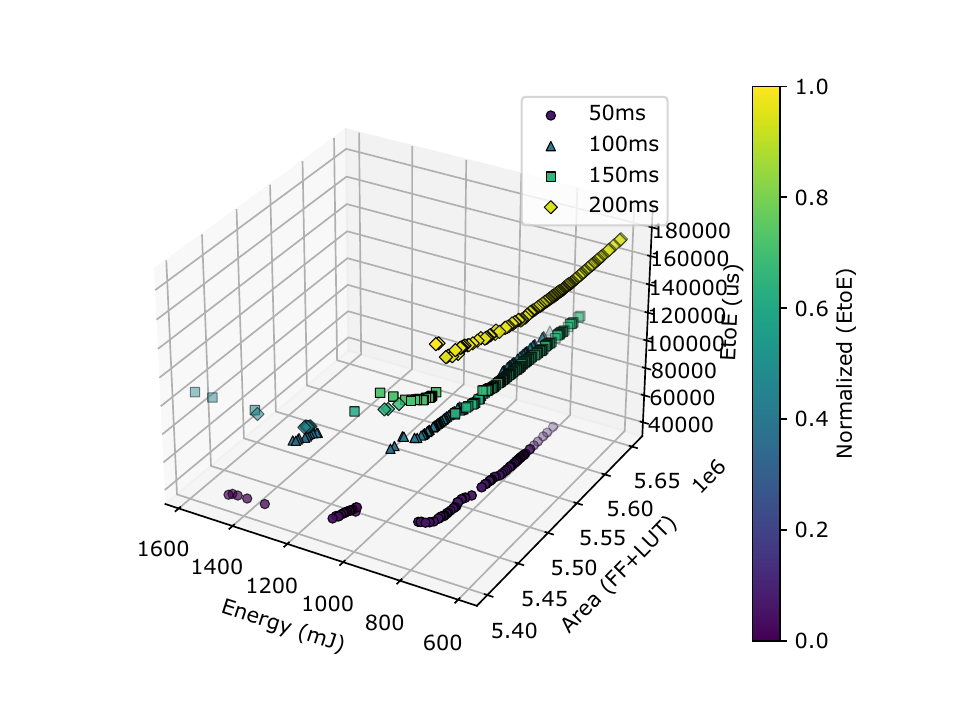}
\vspace{-10pt}
\caption{Visualized 3D comparison of the system-level Pareto-optimal fronts of EtoE-DSE$_{parallel}$ under 50 ms, 100 ms, 150 ms, and 200 ms end-to-end (EtoE) latency constraint. Different level of color in color bar represents the EtoE latency for one of the paths between GPSsensor and Control PSMs}
\label{fig:3D4z_method}
\end{figure}

\begin{figure}[t]
\centering
\includegraphics[width=0.9\columnwidth,keepaspectratio]{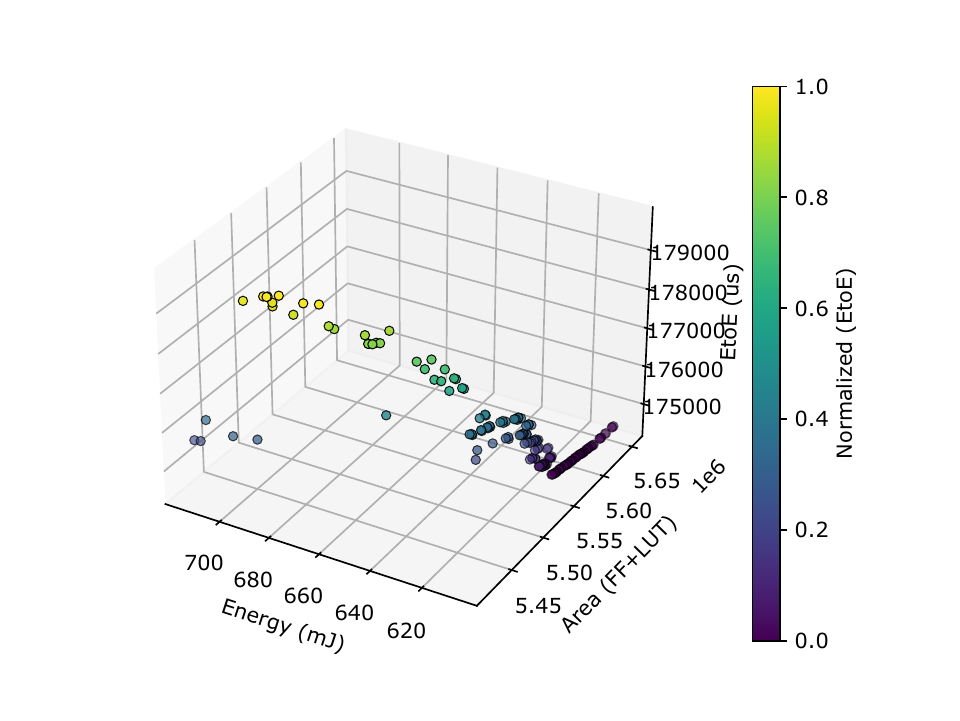}
\vspace{-10pt}
\caption{Visualized 3D system-level Pareto-optimal front of EtoE-DSE$_{parallel}$ under 200ms end-to-end (EtoE) latency constraint. Different level of color in color bar represents the EtoE latency for one of the paths between GPSsensor and Control PSMs}
\label{fig:3D1z_Parallel_200ms}
\end{figure}

\subsection{Design space exploration using EtoE-DSE}
With the assistance of the FDSS algorithm, the design space becomes more manageable, facilitating the identification of Pareto-optimal system-level configurations using LCSO. To evaluate the effectiveness of EtoE-DSE and determine the impact of FDSS, we explored three variants of the proposed algorithm for our experimental system, as discussed in Section \ref{sec:experiments}, in comparison to prior work \cite{gao2021effective}. Fig. \ref{fig:3D5z_50ms} visualizes a 3D comparison of the Pareto-optimal system-level fronts under an EtoE latency constraint of 50 ms for a selected path within two EtoE pairs, resulting from the reference set, prior work (GA), EtoE-DSE$_{unsegmented}$, EtoE-DSE$_{serial}$, and EtoE-DSE$_{parallel}$. Due to the limited space in one figure, we eliminated the prior work (SA), the detailed QoR comparisons are provided later. Given that there are 15 paths for two EtoE pairs, we selected a single path from the EtoE pair (GPS $\gets$ Init to Control $\gets$ Compute) that exhibits the highest EtoE latency to demonstrate the EtoE latency for each design point. The color bar in the figure serves as a visual tool to represent the range of normalized EtoE latency from small to large across a color spectrum. The Pareto-optimal fronts from different methods are distinguished by unique markers. In Fig. \ref{fig:3D5z_50ms}, we observe that, in comparison to prior work and the three variants of EtoE-DSE, the reference set (marked with $*$) encompasses all non-dominated design configurations in terms of energy and area across all latencies. At the same latency level (indicated by the yellow area), the design configurations of prior work (GA) (marked with circles) are much farther from the origin of the x and y axes, indicating higher energy and area compared to the other methods. Furthermore, most of the EtoE-DSE$_{unsegmented}$ configurations (marked with $\triangle$) are also distant from the origin compared with those of EtoE-DSE$_{serial}$ (marked with squares) and EtoE-DSE$_{parallel}$ (marked with diamonds). This observation suggests that the FDSS algorithm effectively segmented and pruned the original ADS system design space, thereby enabling LCSO to more effectively explore system-level Pareto-optimal configurations. However, further comparison of the QoR is necessary to substantiate this observation.

Fig. \ref{fig:3D4z_method} presents a 3D comparison of the system-level Pareto-optimal fronts of EtoE-DSE$_{parallel}$ under EtoE latency constraints of 50 ms, 100 ms, 150 ms, and 200 ms. From the highest constraint of 200 ms to the lowest of 50 ms, we observe an increase in the system's energy consumption. This trend occurs because softer EtoE latency constraints allow for lower operating frequencies for each FSM, MCC, and handshake component to achieve a valid design configuration within the constraint, thereby increasing estimated energy consumption due to longer latency in each component. Notably, a few design points in each estimated system-level Pareto-optimal front exhibit higher energy consumption, indicating that the original design space has more possible design points with higher energy and lower area. As discussed in Section \ref{sec:experiments}, the fitness function prioritizes energy consumption, favoring the exploration of configurations with lower energy relative to area to achieve valid designs under EtoE latency constraints. Fig. \ref{fig:3D1z_Parallel_200ms} visualizes the 3D system-level Pareto-optimal front of EtoE-DSE$_{parallel}$ under a 200 ms EtoE latency constraint, offering an intuitive comparison between each design configuration. This observation—that lower energy consumption corresponds to lower EtoE latency—also applies to this single run of EtoE-DSE$_{parallel}$.

\begin{table}[t]
    \caption{Average Euclidean distance to reference set (AEDRS) score for five runs of vary configurations of EtoE-DSE$_{serial}$, and EtoE-DSE$_{parallel}$ under 50 ms, 100 ms, 150 ms, and 200 ms end-to-end (EtoE) latency constraints.}
    \label{tab:Evaluation_of_configuration}
    \scriptsize
    \centering
    \begin{tabular}{ccccc}
    \hline
     \textbf{Configuration} & \multirow{2}{*}{\textbf{50ms}} & \multirow{2}{*}{\textbf{100ms}}  & \multirow{2}{*}{\textbf{150ms}} & \multirow{2}{*}{\textbf{200ms}}\\ 
     \textbf{(Pop-Gen-Seg)} &  & & & \\ 
    \hline
    \multicolumn{5}{c}{\textit{EtoE-DSE$_{serial}$}}\\  
    \hline
    230-100-44 & 1.63\% & 2.44\% & 3.63\% & 3.89\%\\ 
    
    230-200-22 & 1.24\% & \textbf{1.17\%} & 2.35\% & 1.81\%\\ 
    
    230-400-11 & \textbf{1.15\%} & 1.32\% & \textbf{2.10\%} & \textbf{1.64\%}\\ 
    \hline
    \multicolumn{5}{c}{\textit{EtoE-DSE$_{parallel}$}}\\
    \hline
    230-400-44 & 1.11\% & 0.77\%& 1.49\% & 1.21\%\\ 
    
    230-800-44 & \textbf{0.74\%} & \textbf{0.43\%} & \textbf{0.94\%} & \textbf{1.60\%}\\
    
    230-1600-44 & 0.80\% & 0.63\% & 1.00\% & 2.12\%\\ 
    \hline
    \end{tabular}
    %\vspace{-10pt}
\end{table}

To further evaluate EtoE-DSE, we quantified the QoR for prior work and three variants of EtoE-DSE under different EtoE latency constraints, using the \textit{average Euclidean distance to reference set} (AEDRS), as discussed in Section \ref{sec:experiments}. Before assessing the QoR across different algorithms, we first examined the impact of each hyperparameter in the streamlined GA configuration on QoR, also discussed in Section \ref{sec:experiments}. Table \ref{tab:Evaluation_of_configuration} displays the AEDRS scores for various hyperparameter of the GA configurations of EtoE-DSE$_{serial}$ and EtoE-DSE$_{parallel}$ under EtoE latency constraints of 50 ms, 100 ms, 150 ms, and 200 ms. The lowest AEDRS value for EtoE-DSE$_{serial}$ is observed with the configuration $230-400-11$ (Table \ref{tab:Evaluation_of_configuration}), except under the 100 ms constraint. One primary reason is that if the number of generations is too low, the algorithm will not have sufficient iterations to explore the 44 segmented and pruned subspaces. Thus, for EtoE-DSE$_{serial}$, a balance between the number of generations and segments requires careful consideration, with the number of generations being slightly more critical than the number of segments. For EtoE-DSE$_{parallel}$, the lowest AEDRS score is achieved with $230-800-44$. A larger number of generations does not necessarily guarantee better results. We selected $230-400-11$ for EtoE-DSE$_{serial}$ and $230-800-44$ for EtoE-DSE$_{parallel}$ for subsequent comparisons.

\begin{figure}[t]
\centering
\includegraphics[width=0.7\columnwidth,keepaspectratio]{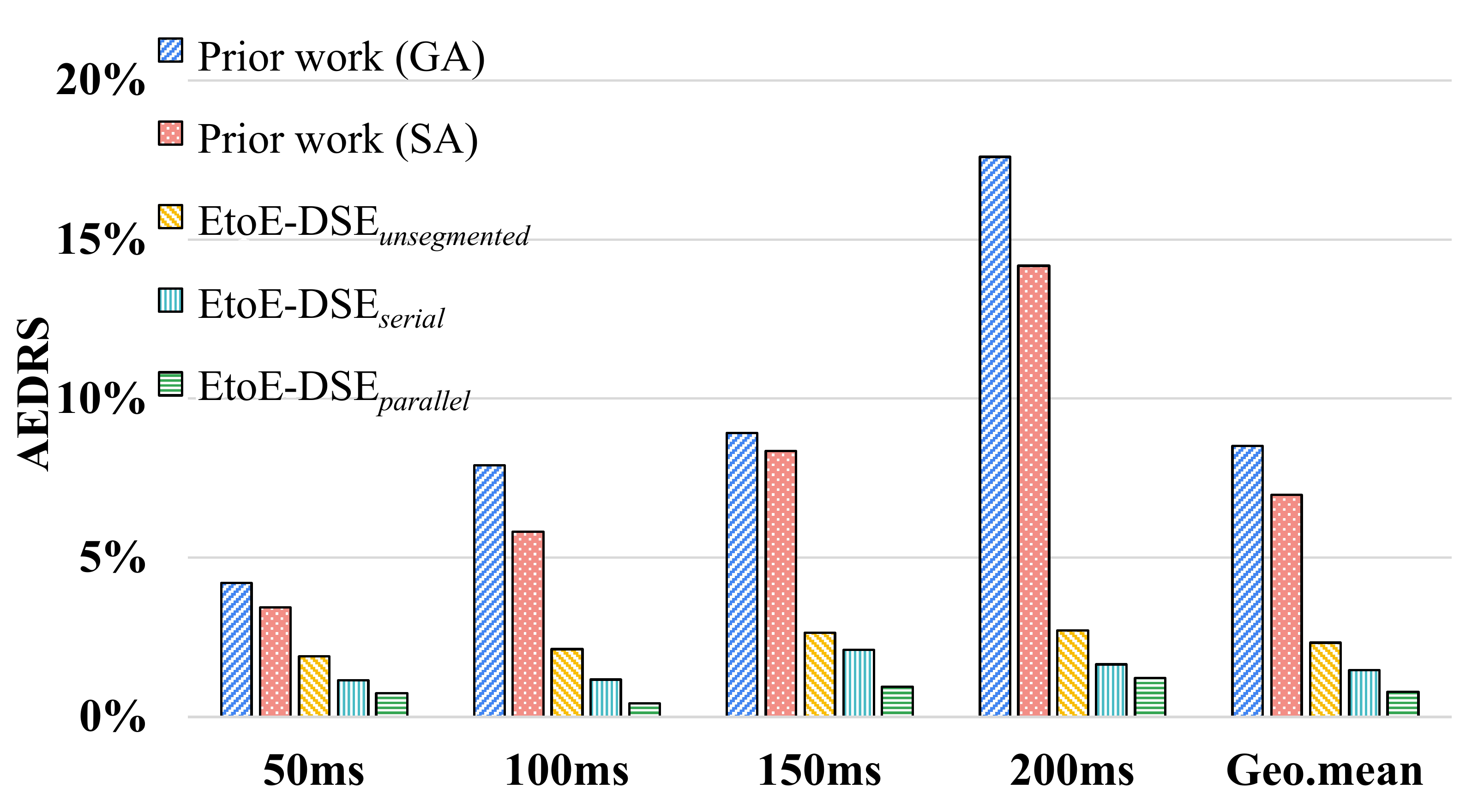}
% \vspace{-10pt}
\caption{Average Euclidean distance to reference set (AEDRS) score from five runs of prior work (GA and SA), EtoE-DSE$_{unsegmented}$, EtoE-DSE$_{serial}$, and EtoE-DSE$_{parallel}$ under 50 ms, 100 ms, 150 ms, and 200 ms EtoE latency constraints.}
\label{fig:ADRS_EtoE}
%\vspace{-10pt}
\end{figure}

Fig. \ref{fig:ADRS_EtoE} shows the average AEDRS score from five runs of prior work (GA and SA), EtoE-DSE$_{unsegmented}$, EtoE-DSE$_{serial}$, and EtoE-DSE$_{parallel}$ under EtoE latency constraints of 50 ms, 100 ms, 150 ms, and 200 ms for the ADS system. Compared to the GA, EtoE-DSE$_{unsegmented}$, EtoE-DSE$_{serial}$, and EtoE-DSE$_{parallel}$ demonstrate an average improvement of 70.02\%, 79.36\%, and 89.26\%, respectively, across all EtoE latency constraints. Although SA outperformed GA by 18.19\%, our work significantly outperformed SA by an average of 63.06\%, 76.39\%, and 84.85\% for EtoE-DSE$_{unsegmented}$, EtoE-DSE$_{serial}$, and EtoE-DSE$_{parallel}$, respectively. These results demonstrate the superiority of our approach to prior DSE approaches.

To evaluate the performance of FDSS, we compared the performance of EtoE-DSE$_{serial}$ and EtoE-DSE$_{parallel}$ (segmented DSE) to EtoE-DSE$_{unsegmented}$. On average, EtoE-DSE$_{serial}$ and EtoE-DSE$_{parallel}$ improved over EtoE-DSE$_{unsegmented}$ by 66.27\% and 82.77\%, respectively, showcasing the benefits of utilizing the FDSS algorithm. During the experiments, under a stringent 50 ms EtoE constraint, half of the segmented and pruned subspace exceeded the threshold value, failing to find any valid design point. Consequently, the subspace and frequency segment were replaced with the original design space and frequency combinations. This hybrid strategy, combining both segmented and pruned subspace with the original design space, ensures that valid design points are not lost due to stringent EtoE latency constraints and enhances the utilization of all available threads in EtoE-DSE$_{parallel}$. Observations from both Table \ref{tab:Evaluation_of_configuration} and Fig. \ref{fig:ADRS_EtoE} indicate that with more relaxed EtoE latency constraints, the AEDRS score tends to worsen for every algorithm. The primary reason is that with softer EtoE constraints, it becomes easier to find design points within the same design space. Comparing EtoE-DSE$_{serial}$ to EtoE-DSE$_{parallel}$, the latter is, on average, 36.5\% more effective, attributed to the use of 2x generation and 44 segments. 

\begin{figure}[t]
\centering
\includegraphics[width=0.7\columnwidth,keepaspectratio]{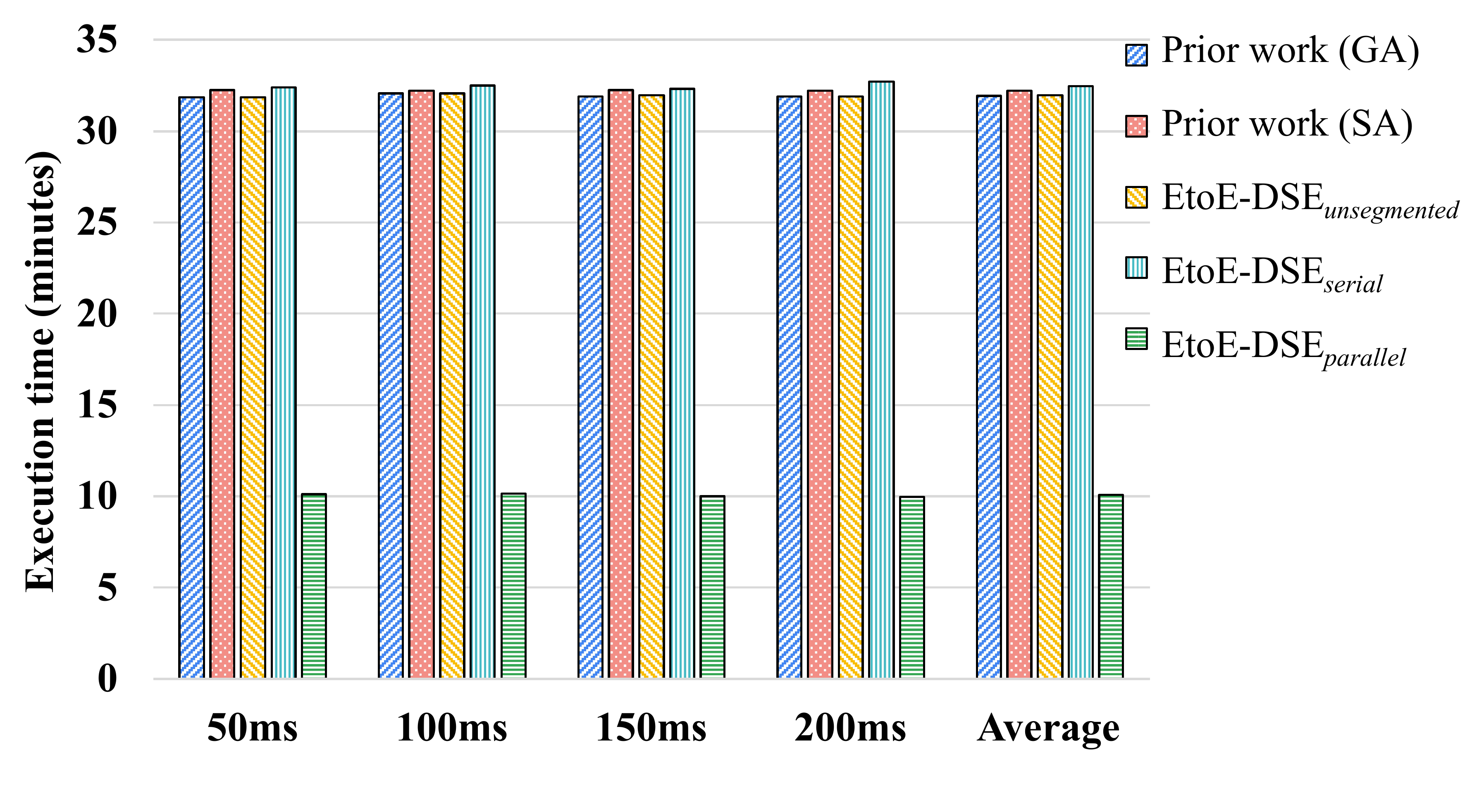}
\caption{Average execution time for five runs of prior work (GA and SA), EtoE-DSE$_{unsegmented}$, EtoE-DSE$_{serial}$, and EtoE-DSE$_{parallel}$.}
\label{fig:executiontime}
\end{figure}

\subsection{Execution time overhead}
We evaluate the overhead of three variants of EtoE-DSE by quantifying the execution time required to identify the system-level Pareto-optimal front, in comparison to prior work (GA and SA) \cite{gao2021effective}. Fig. \ref{fig:executiontime} depicts the average execution time from five runs of GA, SA, EtoE-DSE$_{unsegmented}$, EtoE-DSE$_{serial}$, and EtoE-DSE$_{parallel}$ for the ADS system. GA slightly outperforms SA, EtoE-DSE${unsegmented}$ and EtoE-DSE${serial}$ in execution time by an average of 0.89\%, 0.03\% and 1.59\%, respectively. Conversely, EtoE-DSE$_{parallel}$ \textit{reduces} the execution time by an average of 68.49\% compared to prior work. Overall, these results show that our approach significantly improves the DSE outcomes without introducing much overhead. % The extended execution time observed in EtoE-DSE$_{unsegmented}$ and EtoE-DSE$_{serial}$ represents a trade-off for significantly improved accuracy in finding the system-level Pareto-optimal front.

\section{Conclusion and future work}
We presented a novel holistic approach named \textit{End-to-End Design Space Exploration} (EtoE-DSE). EtoE-DSE efficiently and accurately identifies system-level Pareto-optimal design configurations for multi-component, application-specific embedded systems under both variable timing and EtoE latency constraints. We evaluate the proposed EtoE-DSE approach using a real-world use case of a complex autonomous driving subsystem (ADS) with an original design space of 5.291e+128. Applying various variable timing and EtoE latency constraints to the ADS, targeting FPGA implementation, the FDSS process significantly reduced the design space by several orders of magnitude, without compromising the Quality of Results (QoR). The proposed EtoE-DSE approach rapidly, effectively, and accurately identifies the system-level Pareto-optimal configurations under EtoE constraints, achieving an average QoR improvement by up to 89.26\% compared to prior work that utilized a genetic algorithm for DSE.

A current limitation of the EtoE-DSE is that the EtoE estimation model estimates the worst-case latency. Future work involves developing more sophisticated EtoE latency estimation models that can accurately predict system-level behavior under dynamic operating conditions. We will also further explore trade-offs in the hyperparameter of GA configurations and explore integrating machine learning techniques to improve the accuracy of system-level latency predictions.

\bibliographystyle{IEEEtran}
\bibliography{References}
\balance

\end{document}